\documentclass{elsart}
\usepackage{natbib}
\usepackage{graphicx}
\usepackage{amssymb}
\begin{document}
\begin{frontmatter}
\title{Are Avalanches in Sandpiles a Chaotic Phenomenon?\thanksref{X}}
\author{Maria de Sousa Vieira\thanksref{permaddress}}
\address{International Center for Complex Systems, Departamento de Fisica, UFRN,59072 970 Natal, RN, Brazil.}
\thanks[X]{In honor of Per Bak.}
\thanks[permaddress]{e-mail: mariav\_us@yahoo.com.} 
\begin{abstract}
We show that deterministic systems with strong nonlinearities seem to be 
more appropriate to model sandpiles than stochastic systems or deterministic
systems in which discontinuities are the only nonlinearity. In particular, 
we are able to reproduce the breakdown of Self-Organized Criticality 
found in two well known experiments, that is, a centrally fueled pile
[Held {\sl et al.} {\em Phys. Rev. Lett.} {\bf 65} (1990) 1120] and sand 
in a rotating tray [Bretz {\sl et al.} {\em Phys. Rev. Lett.} 
{\bf 69} (1992) 2431]. By varying the parameters of the  model we 
recover Self-Organized Criticality, in agreement with other experiments 
and other models. We show that chaos plays a fundamental role in the 
dynamics of the system. 
\end{abstract}
\begin{keyword}
Self-Organized Systems; Avalanches; Coupled Map Lattices.\ \ \ \ \ \ \
PACS numbers: 05.65.+b, 45.70.Ht, 05.45.Ra.
\end{keyword}
\end{frontmatter}
In 1987 Bak, Tang and Wiesenfeld showed that certain open, dissipative,  
spatially extended systems spontaneously achieve a critical state  
characterized by power-law distribution of event sizes\cite{soc}.  
They denoted this phenomenon Self-Organized Criticality (SOC) and 
illustrated it using a simple two-dimensional cellular
automaton model for sandpiles.
An enormous amount of
research followed that work. Some natural phenomena that have
been connect to SOC are earthquakes, evolution, interface dynamics,
solar flares, vortices in superconductors, charge density waves among others.
For a review see\cite{review1,review2}. 

The first observation of SOC in granular material was made by 
Frette {\sl et. al}\cite{frette}. They observed that elongated grains of rice 
dropped between two parallel plates with a narrow separation between 
them (that is, in a quasi-one-dimensional geometry) 
present avalanche distribution consistent with SOC. Using more 
symmetric rice a stretched exponential was seen.  

For systems in which the flow of grains is on a two dimensional surface 
different behaviors have been reported. 
To our knowledge, the first of such experiments was performed by 
Jaeger {\sl et al.}\cite{nagel}. They showed that avalanches of 
sand in a rotating drum do not present power-law distribution of event
sizes, instead they found system-spanning avalanches with a narrowly 
peaked distribution. Similar results were found by them for a avalanches 
of sand in a box with an open side\cite{nagel}.   
Using a different set up, Held {\sl et al.}\cite{held} studied avalanche 
distribution by fueling a canonical pile on its apex. They observed that 
if the base of the pile is smaller than a given size, then the
avalanche distribution is consistent with SOC. If the base of the pile 
is increased beyond that critical value, another regime dominated by 
big, nearly periodic avalanches is observed. 
Bretz {\sl et al.} performed experiments 
with granular material in a rotating tray\cite{nori}. They  found big spanning avalanches 
that reset the system, but between those large avalanches a sequence
of small avalanches was found which presents power-law distribution. 

New experiments on granular materials with flow in two dimensions 
have shown 
avalanche distribution  
fully consistent with SOC. The main difference of these experiments with 
respect to the previous ones was a change in the grain type. 
Costello {\sl et al.} studied the avalanche distribution in a canonical 
bead pile\cite{costello} whereas 
Aegerter {\sl et al.} studied a rice pile in a box where grains are dropped 
at one of its edge\cite{aegerter}. In both experiments power-law distribution 
of event size and finite 
size scaling data collapse was seen.  
 
Recently, two simple nonlinear two-dimensional  deterministic models governed 
by coupled maps lattice (CML)
have been introduced by us which reproduce the avalanche distribution observed 
in the experiments with  granular materials\cite{cml2train}. 
We modeled the experiments in 
which a canonical pile is fueled at its apex and the experiment in 
which the granular material is in a rotating tray. 
We observed that in both models there is a region of the parameter 
space where SOC occurs,  
and another region where SOC breaks down. The 
way SOC breaks is different from one model to the other and it is as in 
the experiments. 

Here we expand that work to show that the universality class of those 
systems is not unique. That is, in each model  it changes 
with the parameter values.   
We also study 
the Liapunov exponent of the models and show that    
at the onset of exponential instability 
the distribution of event sizes 
crosses from a power-law one to one dominated by big avalanches. Strong chaos 
is therefore a deciding factor for SOC to be present or not, and  
is closely related to the amount of friction between the grains.  

For the sake of completeness we review the algorithms 
for evolving our systems\cite{cml2train}. We first describe the one 
that governs the centrally fueled pile, called in \cite{cml2train} ``local dropping" (LD): 
(1) Start the system by assigning random initial values for the variables 
$x_{i,j}$, where $x_{i,j}$ is the local slope of the pile,   
so the they are 
below a chosen, fixed,  threshold $x_{th}$.     
(2) Choose a nearly central site  of the lattice  
and update it slope  
according to $x_{i,j}=x_{th}$. 
(3) Check the slope in each element. If an element $i,j$ has $x_{i,j} \ge x_{th}
$, 
update  $x_{i,j}$ according to $x'_{i,j}=\phi(x_{i,j}-x_{th})$, where 
$\phi $ is a given nonlinear function that has two parameters 
$a$ and $d$. 
Increase the slope in all its nearest neighboring element according 
to  $x'_{nn}=x_{nn}+\Delta x/4$, where 
$\Delta x=x_{i,j}-x_{i,j}'$ and $nn$ denotes nearest neighbors.   
(4) If $x'_{i,j} < x_{th}$ for all the elements, go to step (2) (the event, or 
avalanche,  
has finished). Otherwise, go 
to step (3) (the event is still evolving).    
Without losing generality, we can take $x_{th}=1$. 
In our simulation in step (2) we have chosen the site with 
$i=j=L/2$, where $L$ is the lattice size.
The nonlinear function  we use is   
\begin{equation}
\phi (x) = \cases{ 1-d-ax, \ \ \ \  {\rm if} \ \ x < (1-d)/a,\cr 
           0, \ \ \ \  {\rm otherwise.}\cr} 
\label{eq1}
\end{equation}
The parameter $d$ is in the interval $(0 ; 1]$ and is associated with the minimum drop in energy
after an event involving one single element. The parameter    
$a$ is greater or equal to zero and is  
associated with the amount of dynamic friction between the grains. That is,  
the smaller the $a$, the larger the friction and 
the smaller the change in the slope of the pile.   

The algorithm that governs the model used for rotating tray experiment, that is, 
the ``uniform driving case"  (UD),  
is similar to the one given above, with the exception of step (2). 
That step is replaced by:
(2) Find the element in the lattice that has the largest $x$ denoted here 
by $x_{max}$. Then update all the lattice elements according to 
$x_{i,j} \to x_{i,j} + x_{th} - x_{max}$. 

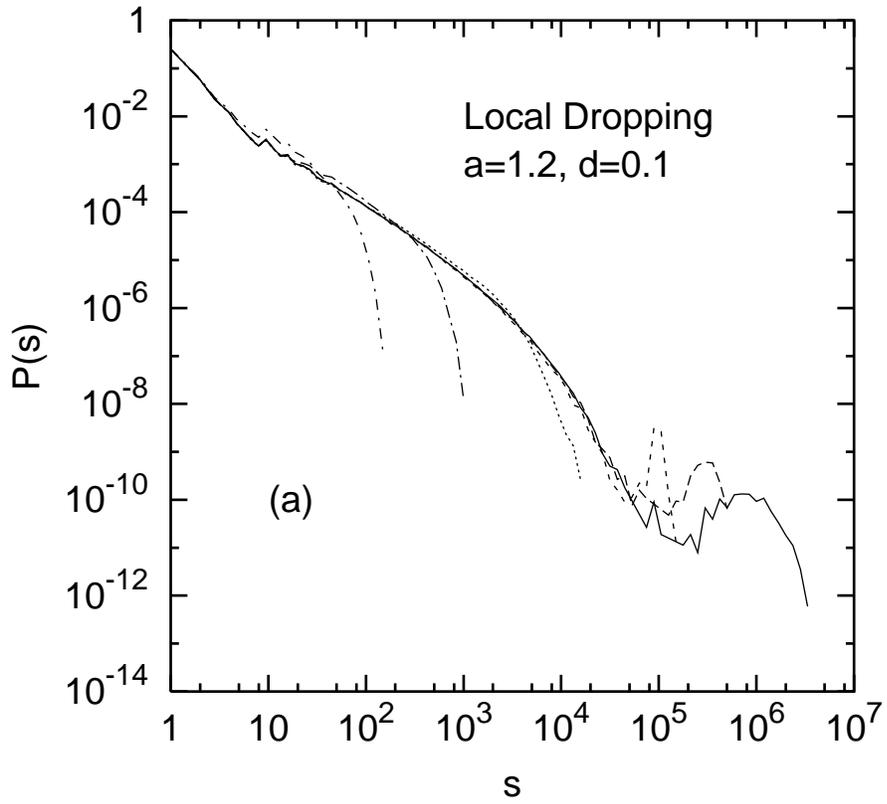
\begin{figure*}
  \vspace{3cm}
\caption{Distribution of event sizes for $a=1.2$, $d=0.1$, $L=8, 16, 32, 64, 128, 256$ for the (a) LD and (b) UD cases.}  
\label{f1}
\end{figure*}

As one evolves the system one can study several quantities, such as 
avalanche size (which is the number of site updates in a avalanche), 
time duration, radius of gyration, etc. Here we limit our study to the 
avalanche size distribution.  We find that  
in both models, when $d=1$ the only nonlinearity in the system 
is a discontinuity between $x_{th}$ and $\phi(0)$  
and we observed that in this case SOC is always present. 
Note that when $d=1$ the UD case reduces to the 
conservative OFC model, which is known to present SOC. 
When $a \le 1$  SOC is found for any 
value of $d$, whereas when  $a>1$ and $d<1$  scaling consistent with 
SOC happens only for $L$ below a given value. For $L$ larger than 
that critical size, big avalanches that belong to a different distribution 
are observed. This is consistent with what is found in experiments with 
granular materials. In Fig.\ref{f1} we show the probability distribution 
of avalanche sizes $P(s)$ as a function of it size $s$ for a regime 
with small $d$ and $a > 1$ ($d=0.1$ and $a=1.2$). 
In (a) we show the case of local dropping for  
$L$ ranging from 8 to 256. In (b) we show similar distribution for 
the uniform driving case for the same system sizes. One sees that for small 
systems sizes ($L \le 32$ in the LD case and  $L \le 16$ in the UD case) 
a scaling regime consistent with SOC is seen. We 
have found finite size scaling data collapse for both cases and for 
the LD case it is reported in \cite{cml2train}. This scaling regime 
reproduces what is observed experimentally in the pile fueled at its 
apex\cite{held}. As the system size is increased a new regime is seen 
where the SOC scaling is lost by the appearance of frequent large avalanches, 
consistent with experiments. Note that for the regime of broken SOC 
the distribution of the small events do not change as the system size 
increases in the LD case, but the the UD case the small avalanches 
become less frequent as the system size increases. This could explain 
why in some experiments, such as \cite{nagel}  modeled by the UD case   
only large avalanches was seen. We have also found that the big
avalanches in the UD case involves all the elements of the system, 
and this does not necessarily occurs in the LD case.      

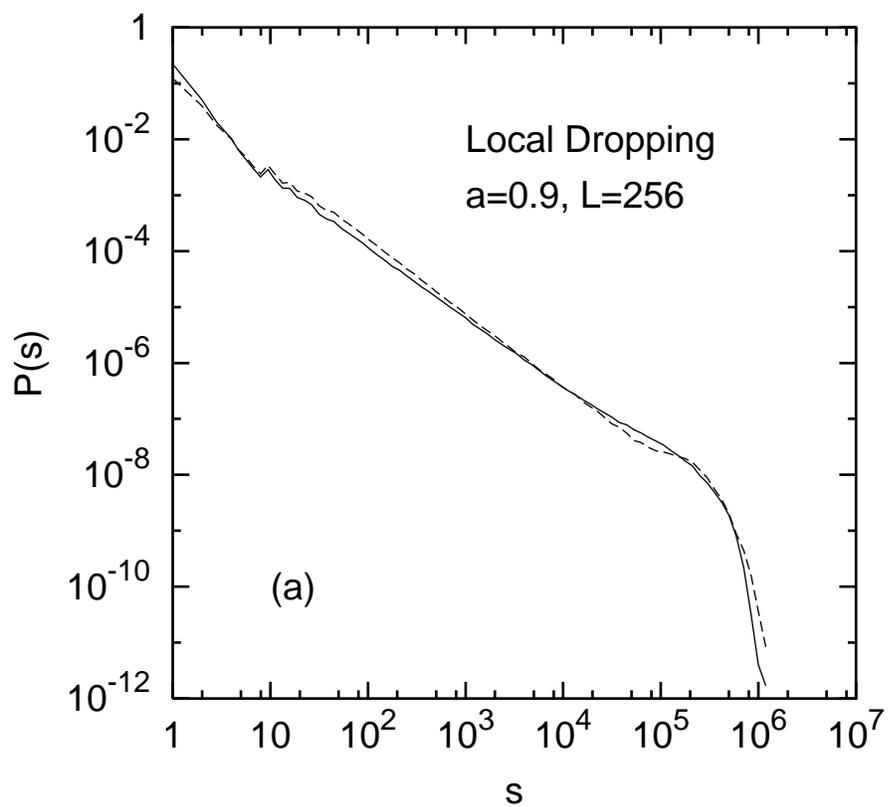
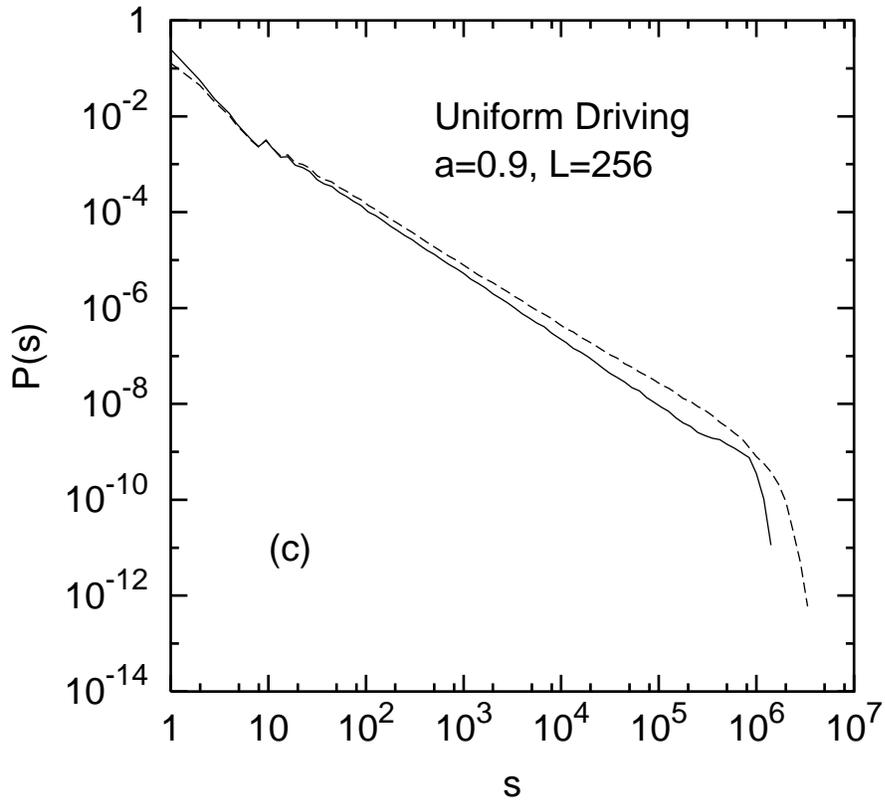
\begin{figure*}
  \vspace{3cm}
\caption{Distribution of event sizes for (a) $a=0.9$, $L=256$ with $d=0.1$ (solid) and  $d = 1$ (dashed) for the LD case. (b) Finite-size scaling data collapse for 
$a=0.9$, $L=64, 128, 256$  with $d=0.1$  (upper set of curves) 
and $d=1$ (lower set of curves) for the LD case. (c) and (d) Same 
as (a) and (b), respectively, except that now 
it applies for the UD case.}  
\label{f2}
\end{figure*}

We now turn our attention to the regime when $a < 1$, where the system 
presents SOC for any value of $d$ and $L$. We plot in Fig.\ref{f2}(a)   
the avalanche distribution for the LD case for two different 
values of $d$, that is, $d=0.1$ (solid line) and $d=1$ (dashed line) with 
$a=0.9$. One sees that $P(s) \sim s^{-\tau}$, characteristic of SOC systems
but the value of $\tau $ is different from one case to the other. We have 
varied the system size and used finite size scaling ansatz  
by plotting 
$P(s)s^{\tau } = s/L^D$ for $L=64, 128, 256$. 
The results are displayed in Fig.\ref{f2}(b).  
The lower 
curves are for $d=1$ and the upper ones for $d=0.1$. We observe that 
finite size scaling is reasonably well obeyed except at the region where the curves 
bend for $d=0.1$. 
We have used $\tau = 1.26$ and $D=2.6$ for $d=0.1$ and 
$\tau = 1.33$ and $D = 2.8$ for $d=1$. We see that the critical 
exponents $\tau $ and $D$ vary as we vary the parameters. Therefore, 
the universality class of this system is not unique.   
In Fig.\ref{f2}(c) we show similar results for the UD case. The solid 
line is for $d=0.1$ and the dashed one for $d=1$.  
In Fig.\ref{f2}(d) we show the data collapse using 
the finite size scaling ansatz for $L=64, 128, 256$. The curves with 
a smoother bending are for $d=1$. 
We have used $\tau = 1.33$ and $D=2.7$ for $d=0.1$ and $\tau = 1.27$ and $D = 3.0$ for $d=1$. Again, we see a change of universality class when the 
parameters of the system are varied.

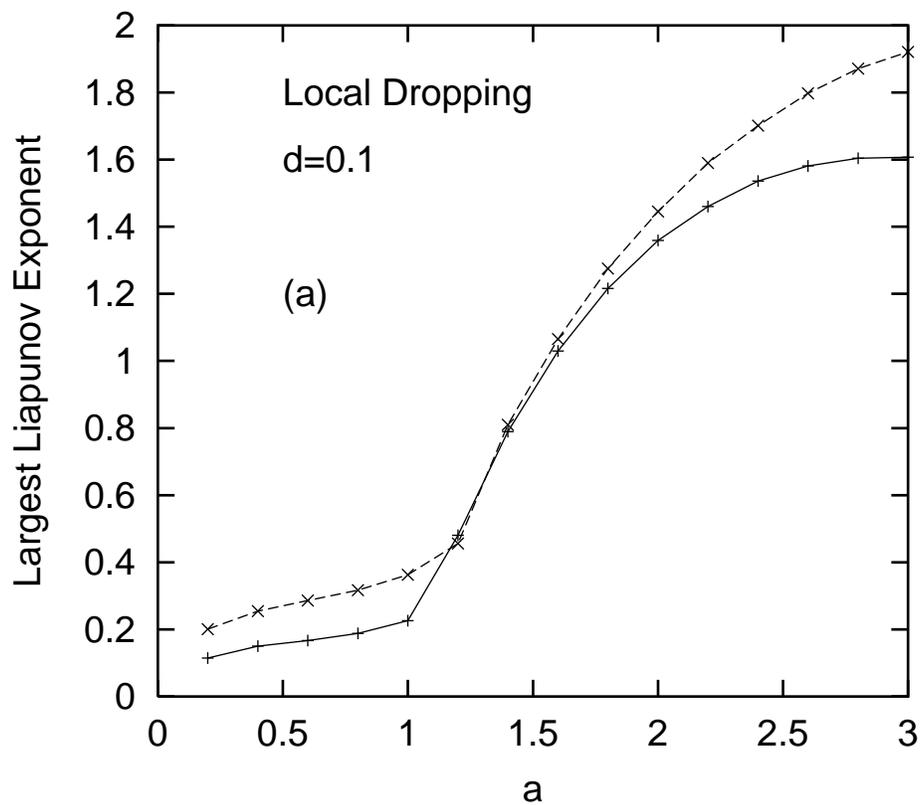
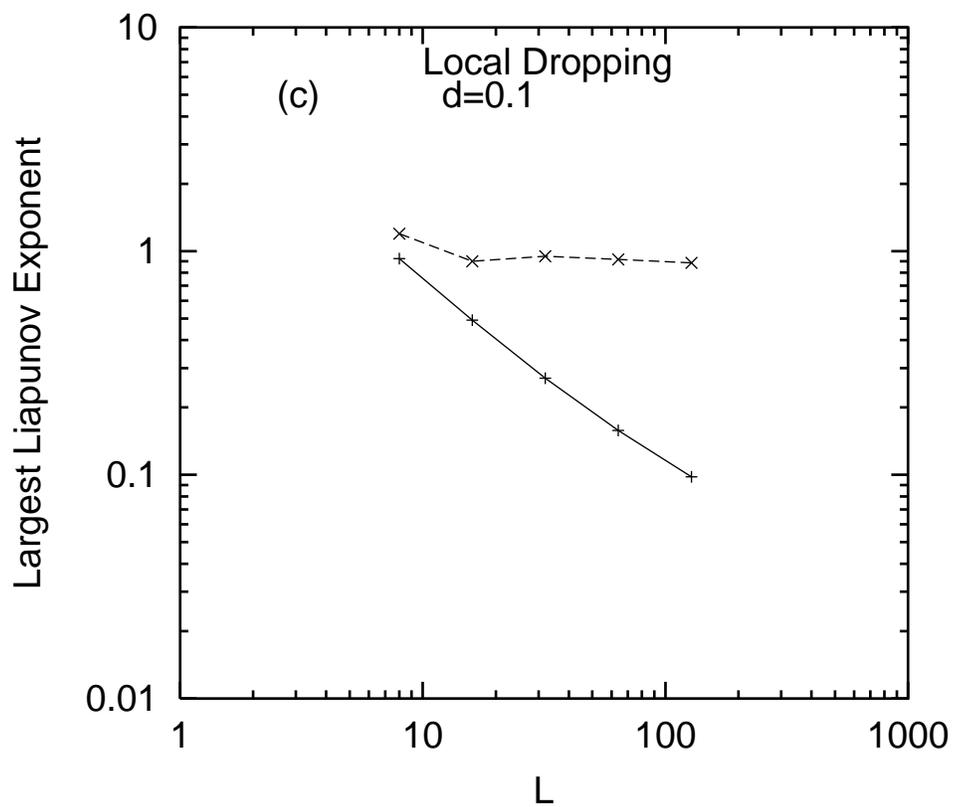
\begin{figure*}
  \vspace{3cm}
\caption{The largest Liapunov exponent as a function of $a$ for $d=0.1$, 
$L=32$ (dashed) and $L=64$ (solid) for (a) LD and (b) UD cases. 
The largest Liapunov exponent as a function of $L$ for $a=0.5$ (solid) and 
$a=1.5$ (dashed) for (c) UD and (d) UD cases.  
}
\label{f3}
\end{figure*}

We have studied the largest Liapunov exponent (LLE) of the system to verify the 
presence or absence of chaos. 
If the LLE is positive, the system is said to be chaotic. 
To study the 
LLE we use the algorithm developed by Benettin {\sl et al.}\cite{liap} and 
the results are displayed in Fig.\ref{f3}.  In (a) and (b) we show 
the LLE as a function of $a$ for the LD and UD cases, respectively.
There we have $d=0.1$, $L=32$ (dashed) and  $L=64$ (solid).  
The figures show that the LLE is positive for any value of $a$ 
and that at $a=1$ a transition happens. If no discontinuities existed
between the $x_{th}$ and  $\phi(0)$ two trajectories with slightly different initial 
conditions would tend to converge when $a < 1$ (see Ref.\cite{cml2train, 
cmltrain}) resulting in a negative LLE. However, because of that 
discontinuity two competing tendencies exist: one for convergence and 
another for separation. However, 
the one for separation is stronger and this results
in a positive LLE, as Fig.\ref{f3} 
shows. This is a weak form of  chaos.  
When  $a>1$ there are no competing tendencies, and the trajectories 
always separate, and this will obviously result in a positive LLE.  
This is a  strong form of  chaos. 
The transition $a<1$ (where SOC is found) to the one in which $a>1$ (where
SOC is broken for large $L$) is markedly different in the two models. 

How the LLE vary with the system size is shown in Fig.\ref{f3}(c) 
for the LD case and Fig.\ref{f3}(d) for the UD case. The dashed line  
refers to $a=1.5$ and the solid one to $a=0.5$. There, $d=0.1$. One 
sees that for small $L$ in both curve the LLE tends to decreases. However, 
as $L$ grows the case in which $a<1$ (where SOC is present for any 
$L$) the LLE continues to get smaller and goes to zero 
(as a power-law function for 
the LD case), whereas when $a>1$ the curves change their behavior 
as $L$ gets bigger (which is consistent with the fact that SOC breaks 
for a given value of $L$ when $a>1$). That is, as the 
system size increases the chaotic behavior 
of the system decreases rapidly in the SOC regime, whereas this does not 
occur in the non SOC regime (in this regime the LLE tends to increase 
or stays approximately constant).  

We summarize our results as follows: the introduction of strong nonlinearities  
in SOC systems causes the appearance of a new behavior as the parameters are changed. This is   
not seen in SOC systems 
in which discontinuities are the only nonlinearity. That is, one is able 
to see the breakdown of SOC as the system size increases, which is seen 
in experiments. Chaos play an important role, since it is only in 
the regime of strong chaos that the breakdown of SOC occurs.

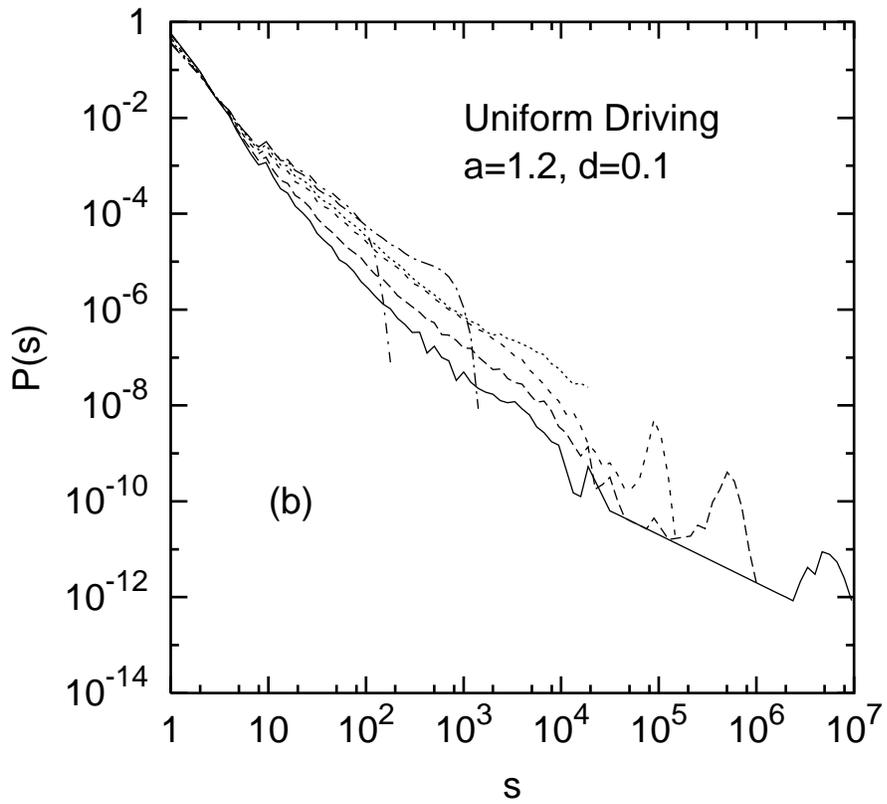
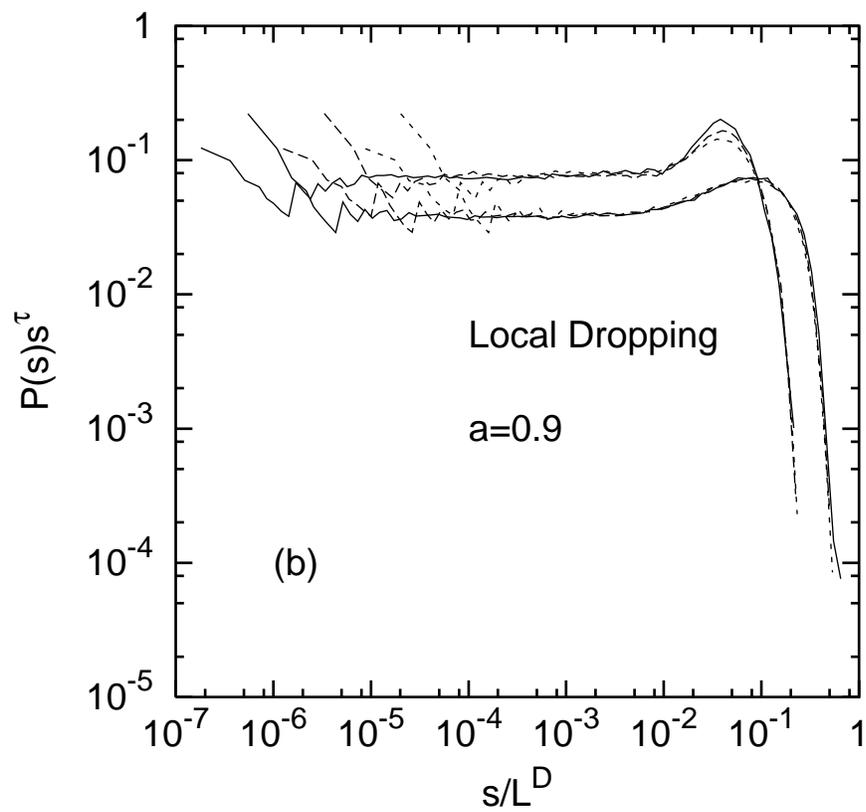
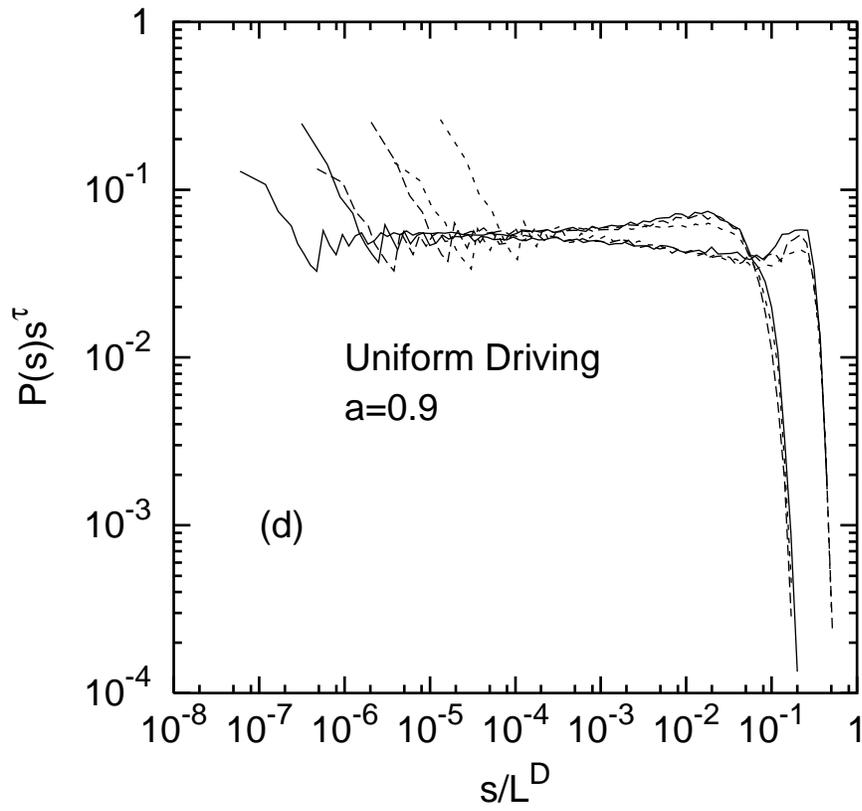
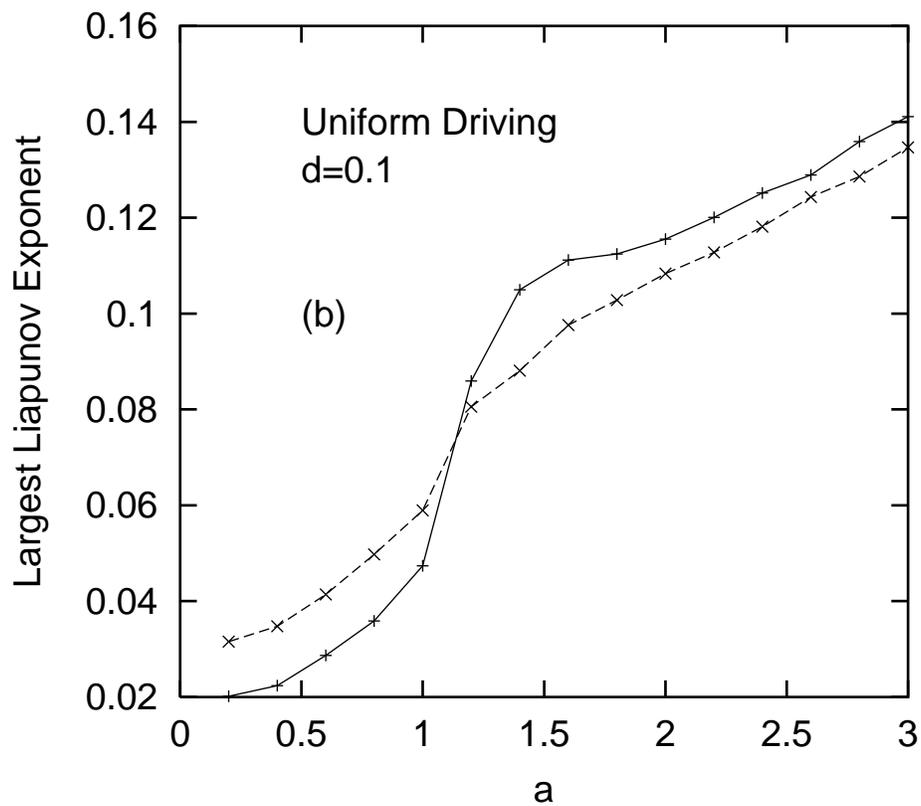
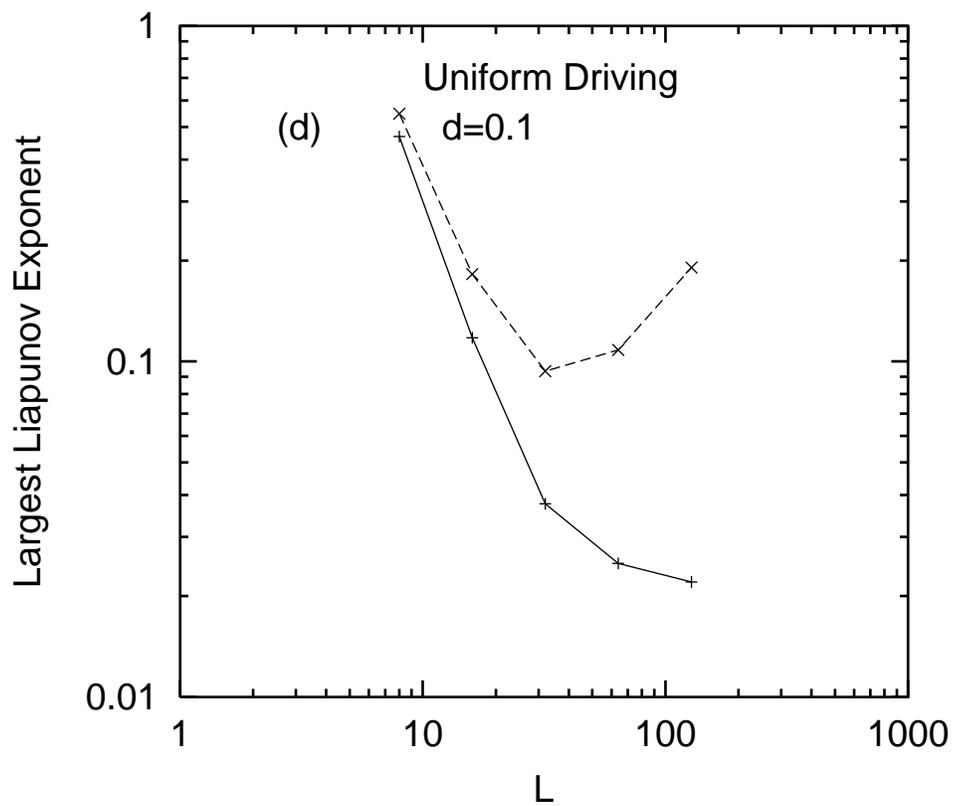
\end{document}